# "With a Thermomix You Lose the Ability to Cook":

# A Kitchen Machine Analogy for Applications of Generative AI in Education

Nikol Rummel[*1, 2], Valentina Nachtigall[*1], and Ernesto Panadero[3, 4]
[*] shared first authorship

[1] Institute of Educational Research,
Ruhr-Universität Bochum, Germany
[2] Center for Advanced Internet Studies, Bochum, Germany
[3] Centre for Assessment Research, Policy and Practice in Education,
Dublin City University, Ireland
[4] Education, Regulated Learning & Assessment group,
Deusto University, Spain

## Abstract

The rapid adoption of generative AI tools such as ChatGPT has sparked intense debate about their risks and opportunities for education, as well as the ways researchers should investigate them. In this paper, we approach these discussions through an analogy with the Thermomix, a smart kitchen appliance that has similarly provoked both enthusiasm and critique. By mapping Thermomix use cases onto examples of learning with generative AI, and situating them within the ICAP and SAMR frameworks, we show how different modes of tool use can either support or undermine meaningful engagement and learning. The Thermomix metaphor underscores that the central question is not whether learners employ AI, but how such use shapes their learning processes. In doing so, we provide a conceptual lens for researchers and practitioners to critically examine – and more effectively guide – the integration of generative AI into educational practice.

## Introduction

When considering the use of generative AI for learning, it may be helpful to reflect on the evolution of kitchen machines. These machines have been around for decades - much longer than generative AI tools have been accessible for learning - and their development and adoption may provide valuable insights into how technological tools can transform established practices without replacing them entirely.

When considering different eating and cooking habits, many would agree that preparing a meal oneself, rather than relying on fast food or pre-packaged convenience items, is not only significantly better for one's physical health but also requires greater effort and creativity. Drawing an analogy to the context of learning with AI one would expect that fully outsourcing school assignments to generative AI tools, without critically engaging with or revising the results, would be equally suboptimal, though in this case, to cognitive development and learning.

However, we argue that this comparison does not have to be a binary choice. Between cooking entirely from scratch and ordering fast food lies a wide continuum of habits, including the use of various kitchen appliances that assist cooking without replacing the cook. The same spectrum exists in learning: between doing all the work unaided vs relying entirely on AI, there are numerous ways AI can support, rather than supplant, meaningful learning. In what follows, we will illustrate this argument through a series of use cases of generative AI tools in learning, drawing analogies to how kitchen machines are used in cooking.

The kitchen metaphor highlights the pitfalls and risks of using AI for learning, but also points to its opportunities for augmenting and enriching learning through AI. By drawing on the metaphor of a manual activity such as cooking – one that demands skill and knowledge while also being emotionally resonant and associated with values such as well-being, active participation, and social connection (see Mosko & Dellach, 2020; Güler & Hasecki, 2021;

Farmer & Cotter, 2021) – this analogy may be particularly well-suited to capturing the multifaceted nature of learning with generative AI.

## The Thermomix: a smart kitchen machine

This paper focuses on the Thermomix, a kitchen appliance developed in Germany that has become a staple in both home kitchens and three-star restaurants (see e.g., Fries et al., 2017). The success of the Thermomix has been compared to that of the iPhone (see Rohwetter, 2015). This smart kitchen appliance - sometimes referred to as a “miracle machine” (translated from Christof, 2024) or a “magic pot”[1] – has a long-standing tradition. It is regularly updated and, in its latest version, offers a wide range of functions including chopping, mixing, kneading, steaming, sous-vide cooking, and fermenting, thereby combining multiple kitchen appliances into one (see e.g., Jarnig, 2024; Daniela, 2021). Beyond the appliance itself, earlier Thermomix models offered an optional recipe chip and a smartphone app, giving access to thousands of recipes that could be sent to the appliance for guided cooking. The app also allowed users to generate shopping lists based on selected recipes and, in some cases, order the required ingredients directly from a specific supermarket (Kollmann, 2018). The latest model, the Thermomix TM7, includes a built-in recipe database accessible via touchscreen, removing the need for recipe chips, though app support remains available. The TM7 is marketed with the slogan “smarter than ever before”[2] and promises to inspire users with new culinary ideas while enabling high-quality, healthy, time-efficient, and stress-free cooking. As the advertised promises already suggest, the Thermomix has the potential to – and in many cases already does – fundamentally transform the way people cook, even replacing traditional methods in some instances.

[1] Translated from: https://www.zaubertopf.de/rezepte/ (Retrieved: September 1, 2025)
[2] See: https://www.vorwerk.com/de/de/thermomix/tm7 (Retrieved September 1, 2025)

Beyond media coverage on its technical capabilities, the Thermomix has also provoked strong cultural reactions. While enthusiasts praise its convenience and versatility, critics question whether relying on it undermines “real” cooking skills. The device has become a social marker, with online forums, blogs, and even media outlets debating whether its guided cooking mode enables or erodes culinary competence. These discussions mirror current controversies around generative AI in education, where optimism about efficiency and access coexists with concerns about authenticity, skill loss, and overreliance. The Thermomix thus provides not only a technological case but also a cultural lens through which to reflect on how new tools reshape practices and identities.

A further noteworthy aspect is the Thermomix’s integration of multiple functions in a single device. Tasks that previously required a separate scale, blender, steamer, or even cookbooks are now concentrated in one machine, supported by an ecosystem of recipes and apps. In this sense, the Thermomix offers more than substitution; it represents a unification of tools that reconfigures the cooking process. Generative AI in learning plays a similar role: it merges functions that were once distributed across search engines, grammar checkers, translators, and note-taking tools into a single interface. This convergence does not simply accelerate existing practices but changes how people approach them, raising questions about which activities are delegated and which remain central to human agency.

**Connecting the Thermomix to models of effective learning and technology use**

With this contextual grounding of the Thermomix as, both, a cultural phenomenon and an integrated tool, we now move to articulate four scenarios that juxtapose statements about this kitchen appliance with what we argue are uses of generative AI in learning that bear analogy. These will be organized in a comparative table to follow, highlighting how each case aligns with the ICAP framework (Chi & Wylie, 2014) and the SAMR model (Puentedura, 2006). The ICAP framework distinguishes four modes of cognitive engagement: Passive, Active,

Constructive, and Interactive. These modes are reflected in learning activities that differ in the extent to which learners invest cognitive effort in modifying existing and creating new learning products and knowledge. Cognitive investment is assumed to be lowest in passive activities, such as listening to a lecture, and highest in interactive activities, in which learners collaboratively co-construct knowledge. Accordingly, these varying levels of engagement are hypothesized to produce different learning outcomes, with higher modes of engagement leading to superior learning outcomes. The SAMR model distinguishes four levels of technology use in education: Substitution, Augmentation, Modification, and Redefinition. These levels differ in the extent to which the use of technology enhances and transforms learning compared to learning without technology. At the lower levels, technology primarily replaces traditional learning activities, such as reading a text in a book, either without functional improvement (Substitution; e.g., reading a text online) or with functional enhancement (Augmentation; e.g., reading an online text that includes links to additional information and materials). At the higher levels, technology use transforms learning by enabling the redesign of existing tasks (Modification; e.g., using texts and multimedia resources for learning) or the creation of entirely new tasks that would not be possible without technology (Redefinition; e.g., visualizing certain aspects of the text). The Thermomix use cases appear not only as a suitable analogy to learning with generative AI, but also, more generally, to learning with technology, as categorized in the SAMR framework, and, even more generally, to learning activities (with or without technology) as distinguished in the ICAP framework. This conceptual lens underscores the suitability of our analogy between cooking with a Thermomix and learning with generative AI.

## Four scenarios of Thermomix and generative AI use

To illustrate the potential risks and opportunities of generative AI for learning, we draw on four scenarios of Thermomix use. Each scenario is paired with a corresponding use of AI in education, thereby highlighting similarities in how technology can both support and undermine the development of skills. Table 1 provides a comparative overview of these scenarios, showing how Thermomix practices map onto uses of generative AI and how each case aligns with the ICAP framework and the SAMR model. In the sections that follow, we elaborate on each scenario, drawing on examples and evidence to illustrate both pitfalls and opportunities.

**Table 1.** Thermomix and generative AI use cases assigned to ICAP and SAMR dimensions

| Thermomix use case | Generative AI use case | ICAP mode | SAMR stage |
|---|---|---|---|
| **#1 "With a Thermomix you lose the ability to cook"** Guided cooking mode, user follows step-by-step instructions without needing prior skills. Risk of skill loss. | Students outsource entire assignments to AI without revision or critical engagement; risk of limited creativity and skill erosion. | **Passive:** learners receive output without active participation. | **Substitution:** AI replaces manual processes (writing, problem solving) with no functional change. |
| **#2 "As a cook I can alter or add ingredients to some extent"** Users modify guided recipes, skip or repeat steps, apply prior knowledge. | Students cross-verify AI outputs with sources, refine prompts, or revise text based on prior knowledge. | **Active:** deliberate manipulations and engagement with materials. | **Augmentation:** technology adds functionality, supporting active adaptation. |
| **#3 "I have become a more creative and versatile cook"** Experienced users employ manual functions for complex dishes, experimenting and broadening repertoire. | Students use AI to brainstorm, outline, evaluate, and build original work, fostering creativity and reflection. | **Constructive:** learners generate new knowledge or products beyond given material. | **Modification:** tasks are redesigned (e.g., assignments prompting critique of AI outputs). |
| **#4 "A new way of cooking"** Emergence of online Thermomix communities, recipe sharing, co-construction of knowledge. | Students interact with AI as partner for dialogue, feedback, adaptive learning, simulating collaborative exchange. | **Interactive:** co-construction through dialogue, feedback, joint reasoning. | **Redefinition:** technology enables tasks previously inconceivable (e.g., real-time adaptive feedback at scale). |

## #1 “With a Thermomix you lose the ability to cook”

### *Thermomix*

A widely popular use scenario for the Thermomix is its *guided cooking mode*. In this mode, users select a recipe – often purchasing specific groceries for it – and then simply follow the step-by-step instructions provided by the device. This functionality has received mixed reactions. Some question whether a Thermomix user can still be considered the cook, or rather an assistant or merely the owner of the device (Schneider, 2019). Others appreciate the guided cooking option, as it enables people to prepare meals who otherwise might lack the skills or confidence to do so. For example, Klünder (2018) interviewed working mothers about their family eating habits, with one noting that her husband could only prepare convenience foods or use the Thermomix, as it was foolproof. This sentiment is echoed in various online forums, where people discuss its utility. Comments include: “Don’t think this is for cooking enthusiasts, but rather for fathers who have no idea about cooking and are now forced to cook sometimes so they don’t have to live on ready meals alone” (translated from DerKaterkatz, 2024) and “I’ve had my Thermomix for a good two years now, and I really thought it through before buying it. One of the main reasons was so that my husband could cook by himself” (translated from Kathyxox, 2019). At the same time, others express concern that such reliance may erode cooking skills, as illustrated by another forum post: “I still hope that future generations will know how to cook properly with pans and other utensils. [...] We love it when our little one stands with us in the kitchen, chopping and cooking with us.” (translated from Baux87, 2019). This guided cooking use case shows how smart kitchen appliances can enable cooking without requiring particular cooking skills, potentially preventing skill acquisition and even leading to skill loss.

### *Generative AI*

Similar concerns have been raised about learning with generative AI, particularly when it is used to complete assignments in a way that *cognitively offloads the entire task*. Critics argue that this reliance risks limiting users' skills, especially for students, by constraining creative thinking, originality, and the development of intuitive abilities (for a review, see Ali et al., 2024). For instance, Hosseini et al. (2023) warned that clinicians relying heavily on ChatGPT-like systems may see their clinical reasoning skills deteriorate, since they are no longer required to actively diagnose and weigh alternatives. In the educational domain, more recently, Radtke and Rummel (2025) showed that novice writers who received AI-produced texts often revised them only superficially, investing less cognitive effort than when revising their own drafts. Together, these examples illustrate how full delegation of learning tasks to AI can diminish engagement with core cognitive processes such as reasoning, creativity, and revision; echoing concerns that the "cook" is no longer really cooking.

### *Passive use and substitution*

Whether for cooking or for learning, such use of smart technologies aligns with the passive mode of cognitive engagement in the ICAP framework (Chi & Wylie, 2014) and the substitution stage of the SAMR framework (Puentedura, 2006). In the ICAP framework, passive engagement, where learners simply receive information without active participation, is assumed to lead to the least learning success, akin to listening to a lecture without taking notes (Chi & Wylie, 2014). Guided cooking in the Thermomix parallels this passive mode, as following instructions step-by-step is less likely to improve cooking abilities. Likewise, when learners completely offload schoolwork to generative AI and accept the results without personal input, they forgo meaningful engagement and fail to develop cognitive skills or subject understanding.

The SAMR framework offers a complementary lens. Substitution, the lowest level of technology integration, refers to replacing older tools without altering functionality (Hamilton et al., 2016). Examples include replacing textbooks with e-books or blackboards with presentation software (Blundell et al., 2022). In guided cooking or AI-generated assignment completion, the technology substitutes the manual processes of cooking or cognitive processes of learning, significantly reducing active engagement.

However, as the examples above suggest, substituting the manual processes of cooking with the Thermomix can also bring about positive functional changes - most notably by enabling people to cook or bake who might otherwise be unable to do so. Transferring these use cases from a smart kitchen appliance to the utilization of generative AI in learning, one could imagine that substituting school assignments with generative AI tools - while unlikely to foster high cognitive engagement or strong learning outcomes - might nonetheless promote inclusion and participation among students who would otherwise struggle to complete their work due to special needs or language barriers.

Another potential opportunity in replacing tasks with generative AI - provided these tasks are low-level and do not require substantial cognitive engagement - lies in freeing up time for more demanding and potentially more productive activities. For example, instead of spending time on grammar and sentence formulation while writing a text, learners can use generative AI to handle these aspects, thereby gaining time for searching, reading, and synthesizing relevant literature, as well as for critically and carefully revising a draft. In this sense, using generative AI for learning in a primarily passive, but still effective way, requires both learners and teachers to reflect on which tasks they wish to delegate to AI and which more active tasks they want to prioritize instead.

Working flexibly, creatively, and with greater time on material originally generated by AI may foster more valuable and productive learning than the substituted task itself. However, this opportunity comes with trade-offs. Reliance on AI for basic functions risks diminishing the acquisition of foundational skills - such as composing grammatically correct texts from scratch - and presupposes a certain level of prior knowledge. Only with this knowledge can learners engage productively, critically, and creatively with AI-generated material. The following use case illustrates such a scenario, in which the passive use of AI to substitute certain tasks is complemented by elaborative higher-level processes.

### #2 "As a cook I can alter or add ingredients to some extent"

#### *Thermomix*

Another use case of the Thermomix involves more active engagement, where users slightly *adapt* the step-by-step instructions provided by the guided cooking mode. In a product test report, the reviewer concludes: "Anyone who knows a little bit about cooking will quickly start to modify the recipes and treat the guided cooking instructions as nothing more than a rough framework. It is practical that you can skip steps, repeat them, or carry them out slightly differently." (translated from Christof, 2024). Similarly, Schneider (2019) cautiously acknowledges in her opinion paper on cooking with the Thermomix: "The cook can alter or add ingredients to some extent, but there is a limit to alteration and the use of one's 'own' recipes." (p. 2). This scenario illustrates that users with a certain level of prior cooking knowledge may not rely exclusively on the appliance's instructions but instead adapt the process according to their own preferences and abilities.

#### *Generative AI*

Drawing an analogy between this mode of Thermomix use and the application of generative AI tools for learning, possible use cases may involve students who cross-verify information provided by an AI tool with credible sources (see Ali et al., 2024), leading them to *adapt or elaborate on the AI-generated output*. As in the Thermomix scenario, such practices require prerequisite knowledge. Specifically, an understanding of AI's limitations, the ability to identify credible sources, and the skills to compare and contrast information across multiple references. Another use case in this category could involve students who, based on their prior knowledge, are dissatisfied with the AI-generated output and refine or clarify their prompts to obtain higher-quality results (see Ali et al., 2024). A further example relates to students revising AI-produced text. As demonstrated in a study by Radtke and Rummel (2025), the more experienced learners are in academic writing, the more thoroughly they revise an AI-generated text. Another example comes from Woo et al. (2025), who implemented a prompt-engineering intervention and found that students who practiced refining prompts showed descriptively higher self-efficacy in using AI tools for academic work and reported significantly greater knowledge of AI concepts. Similarly, Sirnoorkar and Rebello (2025) reported that students preferred feedback generated through prompts designed with structured prompt-engineering techniques over feedback based on their own ad-hoc queries. These findings suggest that, much like cooks who adjust recipes to fit their skills and preferences, learners can engage more actively and productively with generative AI when they refine inputs and critically evaluate outputs.

***Active use and augmentation***

This way of using a Thermomix for cooking, and generative AI for learning, can be linked to the active mode of cognitive engagement in the ICAP framework (Chi & Wylie, 2014) and the augmentation stage of the SAMR framework (Puentedura, 2006). Active learning behaviors - characterized by deliberate motor actions performed with focused attention to manipulate learning materials - are assumed to demand greater cognitive engagement than passive behaviors and, consequently, to lead to greater learning gains (Chi & Wylie, 2014). Examples include taking notes during a lecture, underlining text, or pausing and rewinding a video. Adapting recipes in the Thermomix's guided cooking mode can similarly be described as an active motor manipulation, arising from focused attention to the information and instructions provided. The same principle applies to the use of generative AI in the scenarios described here: rather than passively accepting AI-generated output, learners actively engage with it - by cross-verifying information, refining prompts, or revising the text - thereby requiring and fostering sustained attention.

These technology-assisted cooking and learning examples can also be understood as instances in which manual and cognitive processes are augmented through enhanced functionality provided by technology (see Blundell et al., 2022). For example, when people prepare a dish manually without the aid of a Thermomix, they may see little reason - or have insufficient time - to adapt or modify their traditional recipe. By contrast, when the basic cooking process is taken over by the appliance, experienced users may feel freer to alter the recipe, thereby augmenting their previous cooking habits with greater creativity. A similar dynamic can be observed in AI-assisted learning. When learners possess relevant prior knowledge and remain aware of AI's limitations, generative AI tools used for tasks such as producing text or retrieving information can encourage them to engage in augmented activities they might otherwise devote less attention to - often due to time constraints - such as revising a self-generated text or cross-verifying information from a textbook.

## #3 "I have become a more creative and versatile cook"

### *Thermomix*

The creativity in one's cooking habits can even be enhanced when the Thermomix is used in certain ways. As Christof (2024) notes in his product test report: "The Thermomix can also be used entirely for manual cooking – for reheating, simmering, keeping warm, stirring, kneading dough, steaming, and chopping all kinds of food. The longer you use it and the better you learn to handle it, the more interesting manual cooking becomes." This multifunctionality has also been recognized by three-star chefs and food bloggers, who employ the Thermomix to prepare components of complex and challenging dishes (see Firlus, 2015). The fact that the appliance is used by professional chefs may also persuade skeptics, such as one forum participant who remarked: “As long as I don't see it on a cooking show, I don't think it's necessary” (translated from verreisterNutzer, 2022). With its recipe database, the Thermomix encourages creativity by allowing users to search for recipes in flexible ways. For example, they can enter an event (e.g., Christmas or birthday), select the type of meal (e.g., main dish), or specify certain attributes (e.g., vegetarian). This enables users to discover and experiment with new dishes more easily. Beyond *fostering creativity*, many users praise the Thermomix for making them more versatile cooks. One forum post captures both aspects: “I definitely don't want to be without it anymore. Since then, I've been trying out so many new recipes, making my own spice paste, and generally cooking much healthier meals. I've tried ingredients that I would never have tried before and am cooking very balanced meals” (translated from Rapheli, 2022). The possibility of healthier cooking with the Thermomix is also noted in a small, non-peer-reviewed pilot study by Jarnig (2024), in which a single participant took part in an eight-week intervention program that included preparing healthy dishes with the Thermomix, alongside other measures, and reported improvements in health. These examples suggest that a smart kitchen appliance can influence cooking habits in ways that encourage creativity, versatility, and potentially healthier eating.

### *Generative AI*

In analogy to the use of a Thermomix for fostering creative, versatile, and healthy cooking, generative AI can also be applied in learning contexts to support learners in developing their own ideas from AI-generated outputs. As Kim and Adlof (2024) suggest, tools such as ChatGPT can be leveraged to *promote knowledge construction*, particularly by fostering meaning-making, critical thinking, and self-reflection. To achieve this, teachers would need to adapt their assignments, encouraging students to use ChatGPT’s outputs for activities such as brainstorming new ideas, creating outlines, or evaluating and reflecting on information (Kim &

Adlof, 2024). Recent work has provided concrete illustrations of this potential. Kasneci et al. (2023), for instance, describe how generative AI can act as a “sparring partner” for learners, stimulating critical engagement and helping them to develop and test their own ideas rather than simply consuming information. When used in this way, AI functions less as a source of ready-made answers and more as a catalyst for creativity and reflection.

### *Constructive use and modification*

The scenarios described above for using smart technologies in cooking and learning correspond to the constructive mode of cognitive engagement in the ICAP framework (Chi & Wylie, 2014) and the modification stage of the SAMR framework (Puentedura, 2006). Constructive learning behaviors are generative in nature, involving activities in which learners produce additional or novel outputs that extend beyond the provided learning materials (Chi & Wylie, 2014). This mode aligns with the manual use of a Thermomix - without the guided cooking function - to create new and complex dishes, as well as with the use of generative AI outputs as a basis for brainstorming original ideas, composing independent texts, or drawing critical and personal conclusions.

From the perspective of the SAMR framework, these examples from both cooking and learning illustrate how technology can modify behavior by prompting a redesign of the task itself. In the SAMR framework, the modification stage describes a level at which technology enables a significant redesign of the task rather than merely substituting or augmenting existing processes (Puentedura, 2006). Applied to the Thermomix, this stage is reflected when users move beyond simply replicating familiar recipes in a digital format and instead redesign their cooking process - such as by using the appliance’s manual functions to experiment with complex dishes that would be difficult or time-consuming to prepare with traditional tools. The technology here does not merely make cooking easier; it changes how and what is cooked, opening possibilities for creative experimentation, advanced preparation techniques, and healthier meal designs. For generative AI, the modification stage is evident when teachers redesign their assignments so that learners are not merely tasked with searching for information and synthesizing it, but are instead prompted to engage in new forms of interaction – such as critically evaluating and reflecting on information provided by an AI tool and drawing their own conclusions. In this way, the learning activity is restructured to position AI as a catalyst for meaning-making, critical thinking, and self-reflection, transforming the original task (e.g., searching for and synthesizing information) into one that could not be completed in the same form without the technology, as it would otherwise require more time and multiple sequential learning activities.

## #4 "A new way of cooking"

### *Thermomix*

A final Thermomix use case discussed in this paper is described by Fries et al. (2017) in their case study "Thermomix by Vorwerk – A New Way of Cooking.". They detail how various apps and recipe databases have fostered the development of a Thermomix user community, where individuals *share* self-developed recipes and *receive feedback* from others who have tested them and may suggest adaptations. Fries et al. further note that "users have created one of the world's biggest online culinary communities in which they can look up thousands of recipes created by top chefs or other community members. In forums, users can discuss cooking challenges or find answers to any Thermomix-related cause" (2017, p. 79). This example illustrates how interactivity and feedback emerge when smart, innovative technologies are integrated into established and valued practices – such as cooking – fostering a sense of community and the co-construction of ideas and knowledge.

### *Generative AI*

The core aspects of the said Thermomix use case, namely interactivity and feedback, can also be transferred to the application of generative AI for learning. Specifically, it has been argued that AI tools, such as ChatGPT, can support effective learning and even personalized and adaptive learning due to its interactivity and the provision of real-time feedback (for a review, see Ali et al., 2024). Similarly, Kim and Adlof (2024) argue that ChatGPT could take on the role of different collaboration partners and support learners in their knowledge construction through feedback and conversational interactions in both individual and collaborative learning settings. One example that already illustrates this potential is the paper by Kasneci et al. (2023) who show how ChatGPT can function as a dialogic "sparring partner," engaging learners in interactive exchanges that stimulate debate, critical questioning, and the co-construction of ideas. Further examples refer to the use of generative AI chatbots as a dialogue partner to enhance creative problem solving (Song et al., 2025) or as Socratic friend to promote critical thinking through questioning (Goda et al., 2025). Such use cases move beyond passive or active engagement by situating the learner in an ongoing conversation, where understanding is negotiated and refined in real time.

### *Interactive use and redefinition*

These ways of using the Thermomix and generative AI align with the interactive mode of cognitive engagement in the ICAP framework (Chi & Wylie, 2014) and the redefinition stage of the SAMR framework (Puentedura, 2006). Chi and Wylie (2014) identify interactive learning activities – characterized by constructive dialogues between a learner and a peer, teacher, or computer agent - as the most effective mode for promoting learning. Examples include debating, defending one's position, and asking and answering questions. In this light, discussing cooking challenges or recipes in an online Thermomix community, as well as interacting with ChatGPT to receive feedback on independently developed ideas or solutions, can be viewed as interactive modes of complex cognitive engagement. Such activities are likely to result in highly effective cooking and learning processes and outcomes.

These activities can also be interpreted as a redefinition of tasks and practices through the use of technology (see Hamilton et al., 2016). In the case of the Thermomix, the transformative potential becomes evident when comparing the interactivity within its user community – where members share and discuss new recipes and cooking challenges – with traditional cooking using pans, pots, or a conventional cookbook. In such traditional contexts, family members or friends might occasionally exchange experiences, but it is highly unlikely that, without a smart kitchen appliance and its extensive online community, people would regularly develop new recipes, try out those created by others, request and provide feedback, and engage in sustained discussions about cooking challenges. The Thermomix, in its role as a gateway to such a community, has significantly transformed and redefined the practice of cooking. Similarly, in the context of generative AI, the ability to interact with a responsive partner who can provide real-time feedback and stimulate reflective thinking clearly redefines learning activities. In many school settings, opportunities for truly personalized and adaptive feedback are limited; integrating generative AI makes such individualized learning experiences feasible in ways that would not be achievable without smart technology.

Taken together, these four scenarios illustrate a progression from passive substitution to interactive redefinition, showing how both kitchen technologies and generative AI can either constrain or expand human agency depending on how they are used. In the following conclusion, we draw together these parallels to reflect on what makes someone not just a "user" but a genuine learner and how this metaphor can inform ongoing debates in the learning sciences.

## What makes a cook? Reflections on identity, agency, and evaluation

A central question raised by the analogy between cooking with the Thermomix and learning with generative AI is what actually defines someone as a cook or as a learner. For a cook, is it the ability to recall recipes and techniques from memory, or the capacity to use external aids (e.g. cookbooks, online instructions) to prepare a dish successfully? From this perspective, knowing how to find, interpret, and adapt a recipe may be just as much a marker of culinary expertise as retaining it in memory. In analogy, in the context of learning with AI, a similar debate arises: does learning require students to carry all knowledge internally, or can authentic competence also include the ability to mobilize external resources, including generative AI, to solve problems and construct new understanding? Cox (2024), for instance, describes how AI may shift the identity of learners from makers of their own knowledge to managers of knowledge provided by external devices. Knowledge managers require complex cognitive skills to understand, evaluate, and organize information.

Another dimension concerns the use of tools in creating. Some argue that relying on a Thermomix reduces the agency in cooking, while others see it as simply a different mode of culinary practice, one that still requires judgment, taste and creativity. Likewise, in learning, the use of generative AI can be interpreted either as a threat to originality or as an instrument that, when handled skillfully, supports deeper engagement and more ambitious outcomes. The question, then, is not merely whether a student uses AI, but how that use reflects or undermines their agency (for a scoping review, see Roe & Perkins, 2026), creativity (e.g., Habib et al., 2024), and learning goals (e.g., Pallant et al., 2025).

Finally, being recognized as a cook often depends on the perspective of others. A diner may judge a meal by its taste and presentation without knowing whether it was prepared by hand, with a cookbook, or with a Thermomix. In education, teachers and examiners are in a comparable position: they often evaluate the final product without insight into the extent to which generative AI contributed to its creation. This raises profound questions about authorship, assessment, and trust. For example, Weng et al. (2024) call for assessment methods integrated into students' learning with generative AI to evaluate their critical thinking, communication, and collaboration skills. Just as the diner may be satisfied without asking how the dish was made, educators must reflect on whether evaluation should focus solely on outcomes, or also on the processes by which those outcomes were achieved.

## Conclusion

In this paper, we have explored a domain in which technological innovations – having emerged earlier than generative AI but being likewise oriented toward smart devices and AI-like functionalities – provide a valuable vantage point for reflection. As AI in education the Thermomix has been marked by both exaggerated enthusiasm and sharp criticism, often accompanied by emotionally charged debates. This parallel illustrates a field that shares important similarities with current discussions on generative AI in education, while remaining sufficiently distant to allow for reflection through the lens of analogy. By situating our considerations in this comparative perspective, we contribute a distinct and novel angle to the broader discourse on the risks and opportunities of generative AI in educational contexts.

Our metaphorical exercise serves a broader purpose for both the scientific community and the reader. For researchers, it provides a conceptual lens that makes visible the spectrum of possible uses of generative AI, organized through established frameworks such as ICAP and SAMR. A partly comparable approach to categorizing learning with AI has been proposed by Bauer et al. (2025). They introduce the ISAR model that is interestingly also based on the SAMR and ICAP models to describe the effects of (not only generative) AI compared to learning without AI. By using the kitchen appliance analogy, our goal was to move the debate around the effects of generative AI on learning away from often emotionally loaded stances as we see them in the public media but also in some scholarly articles. As our analogy and use cases demonstrate, the key is not to reduce inquiry to the simplistic question of whether using generative AI tools such as ChatGPT supports learning (for a review of studies that investigate this question, see Deng et al., 2025). Such media-comparison designs risk confounding numerous factors and yield limited insight (see Weidlich et al., 2025). Instead, future research must move toward a deeper understanding of the specific conditions under which generative AI fosters productive learning processes – and when it may hinder them. Only by disentangling these conditions can the field move beyond polarized narratives and toward a nuanced understanding of when, how, and for whom generative AI enhances learning. For practitioners and educators our metaphor offers a language to discuss not only the technical functions of AI but also the deeper questions of identity, agency, and evaluation. More generally, the Thermomix analogy allows readers to reflect critically on how technologies reconfigure human practices, reminding us that the core issue is not whether tools are used, but how they shape the processes and identities of those who engage with them. Ultimately, the challenge is not whether students will “cook” with AI, but what kind of learners and, by analogy, what kind of cooks, they become through its use.

## References

Ali, D., Fatemi, Y., Boskabadi, E., Nikfar, M., Ugwuoke, J., & Ali, H. (2024). ChatGPT in teaching and learning: A systematic review. *Education Sciences*, *14*(6), 643. https://doi.org/10.3390/educsci14060643

Bauer, E., Greiff, S., Graesser, A. C., Scheiter, K., & Sailer, M. (2025). Looking beyond the hype: Understanding the effects of AI on learning. *Educational Psychology Review*, *37*, 45. https://doi.org/10.1007/s10648-025-10020-8

Baux87 (2019, December 8). *Ich hoff trotzdem dass auch die nächsten Generationen noch richtig kochen können mit Pfanne und Co. wir kochen immer frisch* [Comment on the online forum post *Thermomix Ja oder nein*]. Babyforum. Retrieved from: https://www.babyforum.at/discussion/21846/thermomix-ja-oder-nein

Blundell, C. N., Mukherjee, M., & Nykvist, S. (2022). A scoping review of the application of the SAMR model in research. *Computers and Education Open*, *3*, 100093.

Chi, M. T., & Wylie, R. (2014). The ICAP framework: Linking cognitive engagement to active learning outcomes. *Educational psychologist*, *49*(4), 219-243.

Christof, F. (2024, May 11). Thermomix im Test: Kann er einen Skeptiker überzeugen? [Thermomix put to the test: Can it win over a skeptic?] *Futurezone.* Retrived from: https://futurezone.at/produkte/thermomix-test-review-fazit-rezepte-vorwerk-kuechengeraet-smart-home-mixer-kocher/402870341

Cox, G. M. (2024). Artificial Intelligence and the aims of education: Makers, managers, or inforgs?. *Studies in Philosophy and Education*, *43*(1), 15-30.

Daniela (2021, May 5). Thermomix TM6 – Mein Fazit nach 2 Jahren. [Thermomix TM6 – My conclusion after 2 years.] *Die kleine Botin.* [Blog Post] Retrieved from: https://diekleinebotin.at/thermomix-mein-fazit-nach-1-jahr/

Deng, R., M. Jiang, X. Yu, Y. Lu, and S. Liu. 2025. "Does ChatGPT Enhance Student Learning? A Systematic Review and Meta-Analysis of Experimental Studies." *Computers & Education 227*: 105224. https://doi.org/10.1016/j.compedu.2024.105224

DerKaterkatz (2024, May 22). *Ich habe einen hier herumstehen - ist ziemlich überbewertet finde ich. Aber wenn man wirklich keinen Bock auf Kochen hat, kann* [Comment on the online forum post *Lohnt sich ein Thermomix?*]. Gutefrage. Retrieved from: https://www.gutefrage.net/frage/lohnt-sich-ein-thermomix-4

Farmer, N., & Cotter, E. W. (2021). Well-being and cooking behavior: using the positive emotion, engagement, relationships, meaning, and accomplishment (PERMA) model as a theoretical framework. *Frontiers in psychology*, *12*, 560578.

Firlus, T. (2015, October 14). Einsatz in der 3-Sterne-Küche. Wofür braucht man eigentlich einen Thermomix? [Use in 3-star cuisine. What do you actually need a Thermomix for?] WirtschaftsWoche. Retrieved from: https://www.wiwo.de/technologie/gadgets/einsatz-in-der-3-sterne-kueche-wofuer-braucht-man-eigentlich-einen-thermomix/12445260.html

Fries, A., Bergmeister, A. W., & Spindler, M. (2017). Thermomix by Vorwerk–A New Way of Cooking. In: J.-P. Büchler (Eds.). *Fallstudienkompendium Hidden Champions: Innovationen für den Weltmarkt* (pp. 73-90). Wiesbaden: Springer Fachmedien Wiesbaden. https://doi.org/10.1007/978-3-658-17829-1_5

Goda, Y., Arame, M., Toda, M., Handa, J., & Yamada, M. (2025, March). Design and Development of Mondo-GPT: A Generative AI-Integrated Chatbot for Socratic Inquiry and Critical Thinking Enhancement. In *Society for Information Technology & Teacher Education International Conference* (pp. 2753-2760). Association for the Advancement of Computing in Education (AACE).

Güler, O., & Haseki, M. İ. (2021). Positive psychological impacts of cooking during the COVID-19 lockdown period: a qualitative study. *Frontiers in Psychology*, *12*, 635957.

Habib, S., Vogel, T., Anli, X., & Thorne, E. (2024). How does generative artificial intelligence impact student creativity?. *Journal of Creativity*, *34*(1), 100072.

Hamilton, E. R., Rosenberg, J. M., & Akcaoglu, M. (2016). The substitution augmentation modification redefinition (SAMR) model: A critical review and suggestions for its use. *TechTrends*, *60*(5), 433-441.

Hosseini, M., Gao, C. A., Liebovitz, D. M., Carvalho, A. M., Ahmad, F. S., Luo, Y., ... & Kho, A. (2023). An exploratory survey about using ChatGPT in education, healthcare, and research. *Plos one*, *18*(10), e0292216.

Jarnig, G. (2024). *Impact of an Eight-Week Intervention Program With Thermomix Cooking, Healthy Breakfast and Sufficient Daily Physical Activity on the Health Status of an Obese Adult: A Descriptive Pilot Study.* Preprints.org. https://doi.org/10.20944/preprints202408.1651.v1

Kasneci, E., Seßler, K., Küchemann, S., Bannert, M., Dementieva, D., Fischer, F., ... & Kasneci, G. (2023). ChatGPT for good? On opportunities and challenges of large language models for education. *Learning and individual differences*, *103*, 102274.

Kathyxox (2019, December 8*). Ich hab meinen Thermomix nun gut zwei Jahre und habe mir das wirklich gut überlegt vorher. Ein Hauptgrund war dass* [Comment on the online forum post *Thermomix Ja oder nein?!*]. Babyforum. Retrieved from: https://www.babyforum.at/discussion/21846/thermomix-ja-oder-nein

Kim, M., & Adlof, L. (2024). Adapting to the future: ChatGPT as a means for supporting constructivist learning environments. *TechTrends*, *68*(1), 37-46.

Klünder, N. (2018). Zwischen selbst Gekochtem, Thermomix und Schulverpflegung: Innenansichten der Ernährungsversorgung von Familien mit erwerbstätigen Eltern.

[Between home cooking, Thermomix, and school meals: Inside looks at food provision for families with working parents.] *Hauswirtschaft und Wissenschaft*, *11*(8), 1-24.

Kollmann, T. (2018). Die Digitale Transformation in der Küche: Wie der Thermomix das Internet der Dinge in die Küche bringt. [Digital transformation in the kitchen: How the Thermomix brings the Internet of Things into the kitchen]. In: A. Ghadiri, T. Vilgis, T. Bosbach (Eds.). *Wissen schmeckt: Die Magie der Wissenschaften beim Kochen erklärt–mit 16 Rezepten* (pp. 147-167). Wiesbaden: Springer Fachmedien Wiesbaden.

Mosko, J. E., & Delach, M. J. (2021). Cooking, creativity, and well-being: An integration of quantitative and qualitative methods. *The Journal of Creative Behavior*, *55*(2), 348-361.

Pallant, J. L., Blijlevens, J., Campbell, A., & Jopp, R. (2025). Mastering knowledge: the impact of generative AI on student learning outcomes. *Studies in Higher Education*, 1-22.

Puentedura, R. (2006). Transformation, technology, and education [Blog post]. Retrieved from http://hippasus.com/resources/tte/.

Radtke, A., & Rummel, N. (2025). Generative AI in academic writing: Does information on authorship impact learners' revision behavior?. *Computers and Education: Artificial Intelligence*, *8*, 100350.

Rapheli (2022, December 3). Ich habe einen TM6 und ich liebe ihn.

Ich will ihn absolut nicht mehr missen. Ich probiere so viele neue Comment on the online forum post *Thermomix pro und Contra*]. UrbiaCommunity. Retrieved from: https://www.urbia.de/forum/46-haushalt-wohnen/5730201-thermomix-pro-und-contra

Roe, J., & Perkins, M. (2026). Agency in the age of generative AI: a critical review of educational implications. *Academia AI and Applications*, *2*(3). https://www.doi.org/10.20935/AcadAI8422

Rohwetter, M. (2015, October 15). Das iPhone aus Wuppertal. *Die Zeit.* Retrieved from: https://www.zeit.de/2015/42/thermomix-vorwerk-wuppertal-erfolg-kuechenmaschine

Schneider, T. (2019). *Digital eating: #FoodTech and the changing values of eating.* UBVO Opinion Paper Series. Retrieved from: https://www.alexandria.unisg.ch/server/api/core/bitstreams/862a5bf1-4f39-4a6f-9f47-d3144eaf4a2d/content

Song, Y., Huang, L., Zheng, L., Fan, M., & Liu, Z. (2025). Interactions with generative AI chatbots: unveiling dialogic dynamics, students' perceptions, and practical competencies in creative problem-solving. *International Journal of Educational Technology in Higher Education*, *22*(1), 12.

Sirnoorkar, S., & Rebello, N. (2025). *From Self-Crafted to Engineered Prompts: Student Evaluations of AI-Generated Feedback in Introductory Physics.* arXiv preprint arXiv:2508.09825v1. https://arxiv.org/html/2508.09825v1

verreisterNutzer (2022, June 26). *Solange ich das Teil nicht in irgendeiner Kochsendung sehe glaube ich nicht dass man es braucht.* [Comment on the online forum post *Vorwerk thermomix?*]. GuteFrage. Retrieved from: https://www.gutefrage.net/frage/vorwerk-thermomix

Weidlich, J., Gasevic, D., Drachsler, H., & Kirschner, P. (2025). ChatGPT in education: An effect in search of a cause. *Journal of Computer Assisted Learning 41*(5), e70105. https://doi.org/10.1111/jcal.70105

Weng, X., Qi, X. I. A., Gu, M., Rajaram, K., & Chiu, T. K. (2024). Assessment and learning outcomes for generative AI in higher education: A scoping review on current research status and trends. *Australasian Journal of Educational Technology*.

Woo, D. J., Wang, D., Yung, T., & Guo, K. (2026). Effects of a prompt engineering intervention on undergraduate students' AI self-efficacy, AI knowledge and prompt engineering ability: A mixed methods study. *British Educational Research Journal*, *52*(2), 1442-1469. https://doi.org/10.1002/berj.70087